# scientific reports

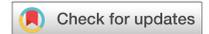

OPEN

# Generation of giga-electron-volt proton beams by micronozzle acceleration

M. Murakami[1✉], D. Balusu[1,2], S. Maruyama[1], Y. Murakami[1] & B. Ramakrishna[2]

Our proposed ion acceleration scheme, micronozzle acceleration (MNA), generates proton beams with extremely high kinetic energies on the giga-electron-volt (GeV) order. The underlying physics and performance of MNA are studied with two-dimensional particle-in-cell simulations. In MNA targets, a micron-sized hydrogen rod is embedded inside a hollow micronozzle. Subsequent illumination of the target along the symmetric axis by an ultraintense ultrashort laser pulse forms a strong electrostatic field with a long lifetime and an extensive space around the downstream tail of the nozzle. The electric field significantly amplifies the kinetic energies of the accelerated protons, and $\gtrsim$ GeV protons are generated at an applied laser intensity of $10^{22}$ W/cm$^2$.

Laser-plasma-based ion acceleration displays unique properties such as a high directionality and laminar flow[1], spatial confinement on the micrometer order ($\sim \mu$m) and temporal compactness ($\lesssim$ ps), containing up to $10^{13}$ particles in a pulse width. These properties hold promise for a wide range of applications, including diagnostic tools in proton radiography experiments[2,3], compact particle accelerators[4–6], creation of high-energy-density (HED) matter[7], proton fast ignition[8], and injection into conventional accelerators[9]. For example, high-quality proton beams with energies $\gtrsim$ 200–300 MeV are required in hadron therapy applications to shrink tumor cells[10,11].

Various acceleration mechanisms have been developed in the past few decades to realize high-quality and high-energy ion beams. Proposed mechanisms include target normal sheath acceleration (TNSA)[12–14], radiation pressure acceleration (RPA)[15–17], collisionless shock acceleration (CSA)[18,19], break-out afterburner (BOA)[20,21], magnetic vortex acceleration (MVA)[22,23], and coulomb explosion (CE)[24–26]. Among these, TNSA has been extensively investigated in simulations and experiments due to its ease of implementation[27].

Still, future potential applications with higher robustness and feasibility necessitate to achieve even higher kinetic energies. For example, laser pulse absorption by electrons must be maximized to enhance the transfer of laser energy to ions. One effective approach involves engineered foil targets with a structured design in the primary laser interaction region, which is a departure from using flat foils. The success of the structured targets in enhancing the conversion efficiency and temperature for the laser-driven electrons is noteworthy, as demonstrated in both particle-in-cell (PIC) simulation and experimental results. Targets include cone targets[28–30], double-layer targets[31,32], foams[33,34], carbon nanotubes[6,35], nanowires[36–38], microbubbles[39,40], and microtubes[41–43]. In addition, several other structured targets[44–47] have been proposed to enhance the kinetic energy of the ion beams.

In recent years, different schemes have been proposed to achieve GeV-class proton energies[15,20,48–52] by applying ultrahigh laser intensities of the order of $10^{22}$ W/cm$^2$. Herein we propose a new approach in the pursuit of achieving GeV protons. Although our approach also uses high laser intensities of $\sim 10^{22}$ W/cm$^2$, the underlying principle significantly differs from other previously proposed schemes.

Here, we propose a novel ion acceleration scheme, which we referred to as micronozzle acceleration (MNA). MNA is achieved through a unique target structure. Figure 1 shows a cross-sectional schematic view of the MNA target on the $x$-$y$ plane. The MNA target employs a micronozzle, which houses a solid hydrogen rod (H-rod). The H-rod is placed at around the nozzle neck to maximize the proton emission. As an example, this study employs aluminum as the nozzle material. The micronozzle functions like a power lens, such that laser energy absorbed by the nozzle leads to the generation of strong electric fields to accelerate the protons contained in the center-located H-rod.

The micronozzle facilitates a three-stage ion acceleration process resulting from the unique structure. In the first stage, irradiating with an ultraintense laser pulse produces hot electrons with relativistic energies of mega-electron-volts (MeV) on the inner surface of the nozzle head. The electrons flow on and beside the H-rod in a converging manner due to the tapered nozzle-head structure. The negative electron cloud in the nozzle cavity and the positively charged inner wall surface of the nozzle further accelerate the protons emitted from the H-rod toward the nozzle exit.

[1]Institute of Laser Engineering, Osaka University, 2-6 Yamadaoka, 565-0871 Suita, Osaka, Japan. [2]Department of Physics, Indian Institute Technology Hyderabad, Sangareddy, Telangana 502285, India. ✉email: murakami.masakatsu.ile@osaka-u.ac.jp

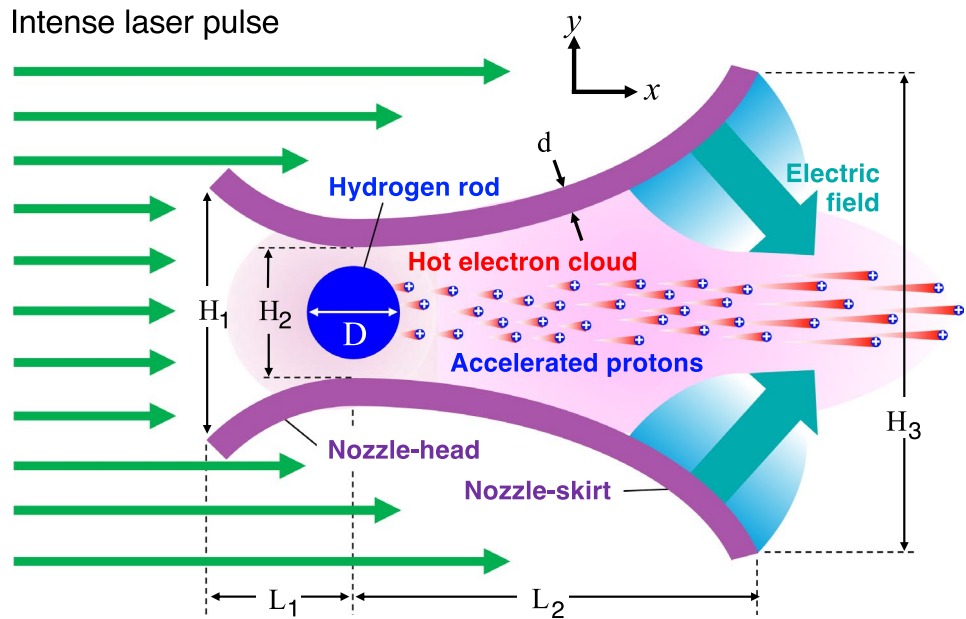

**Fig. 1**. A cross-sectional schematic view of MNA target. MNA target employs a micronozzle housing a solid hydrogen rod (H-rod), which is placed at around the nozzle-neck to maximize the proton emission. Aluminum is here employed as the nozzle material just as an example. The role of the micronozzle can be understood as a kind of power lens, as it were, to integrate the applied laser energy onto the tiny H-rod to bring about significantly higher energies than without the nozzle structure.

Meanwhile, the peripheral part of the incident laser pulse missing the nozzle head irradiates the outer surface of the nozzle skirt, generating another bunch of hot electrons. These hot electrons flow partly into the downstream vacuum and partly back into the nozzle cavity. Because of a large amount of hot electrons emission, the nozzle tail is strongly charged up in positive. In the second stage, protons exiting the nozzle are further accelerated by the electrostatic field generated on the inner surface of the nozzle skirt. This second stage acceleration is more effective than the first one due to the long distances ranging over the nozzle aperture scale. Finally, in the third stage, the protons keep their acceleration gaining a substantial amount of thermal energy transferred from hot electrons during free expansion even after the laser illumination.

For instance, illuminating a MNA target with a peak laser intensity of $I_L = 3 \times 10^{21}$ W/cm$^2$, which is the standard intensity attained in laboratories equipped with modern high-power femtosecond lasers, gives rise to a maximum accelerated proton energy $\mathcal{E}_p \geq 400$ MeV. Such an energy is typically 3 - 4 times higher than those obtained via the TNSA scheme in use of simple planar targets. Moreover, at an intensity of $I_L = 1 \times 10^{22}$ W/cm$^2$, which is the highest attainable level in current state-of-the-art laser facilities, it is numerically demonstrated that the maximum proton energy exceeds $\mathcal{E}_p \sim 1$ GeV.

## Results
### Two-dimensional PIC simulations

We perform two-dimensional (2D) PIC simulations using the open-source fully relativistic code EPOCH.[53] In the 2D ($x$, $y$) PIC simulations, the system is assumed to be uniformly long-stretched along the $z$-axis. The curves of the inner surfaces of the nozzle head and nozzle skirt are designed using parts of a circle and an ellipse, respectively. The nozzle head and nozzle skirt are then smoothly joined on the nozzle-neck boundary. The boundary corresponds to the vertical dashed line in Fig. 1 passing through the H-rod center. In practice, the resultant MNA target structure can be specified by the following scale parameters: the vertical lengths of the nozzle ($H_1$, $H_2$, and $H_3$), the horizontal lengths ($L_1$ and $L_2$), the H-rod diameter ($D$), and the nozzle wall thickness ($d$). The default settings in the following simulations are $H_1 = 5.3$μm, $H_2 = 2.8$μm, $H_3 = 12.0$μm, $L_1 = 3.1$μm, $L_2 = 9.9$μm, $D = 2.0$μm, and $d = 0.6$μm. This paper focuses on clarifying the underlying physics of MNA in terms of the default parameters. Therefore it is not our main purpose in this paper to optimize the set of parameters for the target structure and the laser pulse. The simulation box size placed on the $x$-$y$ plane is 100 – 200μm (along the $x$-axis, depending on the applied laser intensity) ×40μm (along the $y$-axis) at a rate of 100 cells/μm or equivalently 10 nm/cell. The simulation box size, (100 – 200)μm×40μm, is set such that significant amount of charged particles are not lost out of the simulation box to affect accelerated proton dynamics with energies of ∼ GeV.

From the left boundary of the simulation box, a p-polarized laser pulse (the laser electric field oscillates along the y-axis) with the laser wavelength $\lambda_L = 0.8\,\mu m$ is irradiated along the x-axis. The other three sides of the simulation box are treated as open boundaries. The applied laser pulse has a Gaussian shape both temporally with a pulse width of $\tau_L = 100$ fs (FWHM: full width at half maximum) and spatially with a spotsize of 10 μm (FWHM) along the y-axis. Note that laser peak time $t_p$ is set such that $t_p = 1.5\tau_L = 150$ fs. The Al-nozzle and the H-rod are constructed of solid aluminum and solid hydrogen with atomic number densities of $6 \times 10^{22}$ cm$^{-3}$ and $5 \times 10^{22}$ cm$^{-3}$, respectively. We assume fully ionized states for the Al-nozzle with $Z = 13$ and the H-rod with $Z = 1$. Each square cell for these materials is filled with 100 pseudo ions and 200 pseudo electrons.

### Run-up and main-drive phases

When the laser hits the target, the central part of the incident laser reflectively focuses on the H-rod due to the ramp structure of the nozzle neck, while laser absorption generates hot electrons on the inner surface. Figures 2a, b show the lateral component of the electric field $E_y$ (compare Fig. 1) distributions at $t = 110$ fs without and with the Al-nozzle, respectively, when the peak $I_L = 3 \times 10^{21}$ W/cm$^2$. The maximum values of $E_y$ on the H-rod surface are $E_y = 1.5 \times 10^{14}$ V/m in Fig. 2a but is amplified to $E_y = 4.0 \times 10^{14}$ V/m in Fig. 2b. This corresponds to an amplification factor of 2.5 for the $E_y$ and 6.2 for the $I_L \sim E_y^2$. Thus, the nozzle-head structure substantially increases the energy influx of the laser light onto the H-rod. The H-rod emits protons from the surface as it is heated by the laser light and hot electrons. The emitted protons begin to expand along the x-axis. In addition, the part of the incident laser that misses the nozzle neck falls on the outer surface of the nozzle skirt. It also generates hot electrons flowing into the vacuum inside the nozzle. The charge separation produces strong electric fields, especially in the peripheral region of the nozzle skirt.

Figure 3a–d show snapshots of the longitudinal component of the electric field $E_x$ (compare Fig. 1) at different times, $t = 50, 100, 150$ (laser peak time), and 200 fs, with $I_L = 1 \times 10^{22}$ W/cm$^2$. For $t \lesssim 100$ fs, strong electric fields on the order of $E_x \approx 10^{14}$ V/m appear only in the vicinity of the inner surface of the nozzle. However, after the laser peak ($t \gtrsim 150$ fs), the strong electric fields spread over a significantly wider area around the nozzle exit. These long-range electric fields are formed with the characteristic structure of the micronozzle but not formed with planar targets.

Figure 3e–h show snapshots of the proton density plotted coherently with the electric fields in Fig. 3a–d, respectively. Initially, the sheath electric field generated on the H-rod surface accelerates the protons. The protons are gradually accelerated in the nozzle-skirt volume (run-up phase). Once they enter the strong electric field generated around the nozzle exit, the protons undergo strong acceleration (main-drive phase).

Figure 3i–ℓ show snapshots of the electron density in logarithmic scale. While most of the electrons are relatively cold and therefore kept in the nozzle (the red zones), substantial amount of hot electrons with energies > MeV are emitted into vacuum to form the electron clouds depicted by the blue and black zones. It should be noted that the electron densities inside and periphery of the nozzle are kept significantly high; those electrons play an important role in the run-up and the main-drive phases.

Figure 3m–p show snapshots of the magnetic fields in z-direction, $B_z$. Along the inner surface of the nozzle, $B_z = 200 - 300$ kT are observed. As briefly elaborated in Fig. 3h, the protons split into the two branched flows - one along the x-axis accelerated and also well collimated by the electric force, and the other deviating from the x-axis under the v × B force.

Figure 4a plots the proton energy spectrum $dN/d\mathcal{E}_p$ obtained by the MNA scheme along with two different targets (H-rod and foil) to compare with the MNA results. The H-rod ($2\mu m$-diameter) is just the bullet body taken out from the MNA target, while the foil target is composed of a 1μm-thick aluminum with 50nm-thick solid hydrogen layer over-coated on the rear surface. In principle, the TNSA scheme explains the proton accelerations of the H-rod and the foil target. Here it is noteworthy that the proton bunch ejected out of the nozzle exit is subject to the relatively uniform electric fields through the main-drive and afterburner phases. This results in the leveled-off energy spectrum for $400 \lesssim \mathcal{E}_p \lesssim 800$ MeV instead of the monotonically decaying shapes seen in the H-rod and the foil targets, as can be observed in Fig. 4a. It should be also noted that in such a high $I_L$ region where the electron Debye length is much greater than the H-rod radius, the protons emitted from the H-rod behave like a Coulomb explosion.

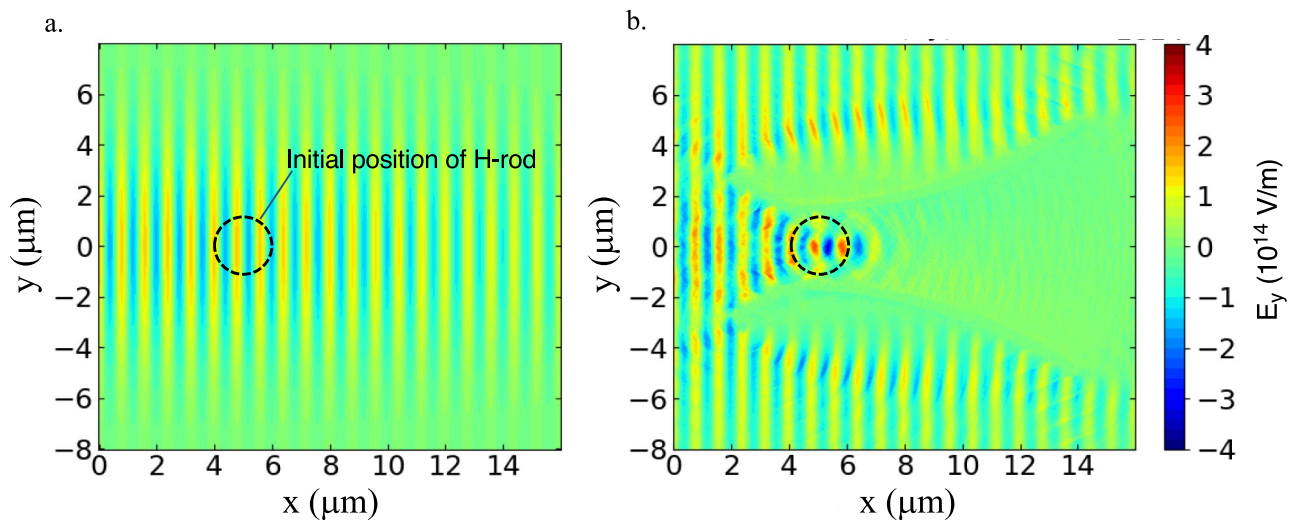

**Fig. 2.** Distributions of the lateral electric fields $E_y$ (compare Fig. 1) at t = 110 fs without (left) and with (right) the nozzle, respectively, where the peak laser intensity of $I_L = 3 \times 10^{21}$ W/cm$^2$, the pulse width of $\tau_L = 100$ fs, and the spot size of 10 μm (FWHM) are employed. The nozzle-head structure works to amplify the energy fluxes of both laser light and hot electrons in terms of the nozzle-head.

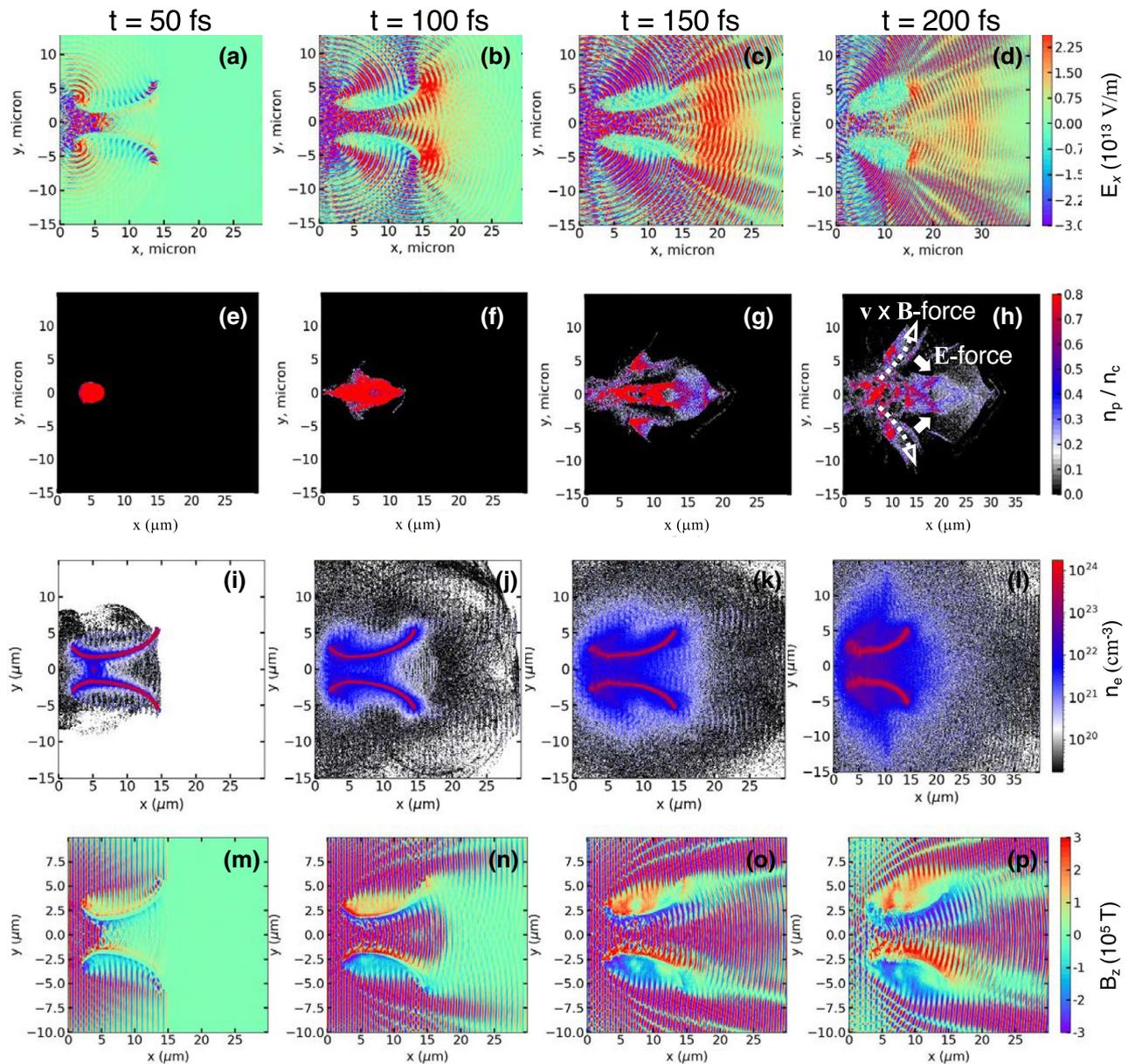

**Fig. 3.** 1st row (**a**)–(**d**): 2D profiles of the longitudinal electric fields $E_x$ ( compare Fig. 1). 2nd row (**e**)–(**h**): proton density profiles at different times, $t$ = 50, 100,150 (laser peak time), and 200 fs for the MNA target; $n_c$ denotes the critical density. The applied laser intensity and the pulse width are $I_L = 1 \times 10^{22}$ W/cm$^2$ and $\tau_L = 100$ fs, respectively. 3rd row (**i**)–(**l**): electron density in log scale. 4th row (**m**)–(**p**): magnetic field in the z-direction.

Figure 4b plots the temporal evolution of the maximum proton energy $\mathcal{E}_{max}$ for the three targets in Fig. 4a. At $t = 1$ ps, $\mathcal{E}_{max} \simeq 900$ MeV for MNA is appreciably higher than those of the H-rod and the foil targets (TNSA), $\mathcal{E}_{max} \simeq 380$ MeV, with an absolute difference of $\simeq 500$ MeV and a factor of $\mathcal{E}_{max}$(MNA)/$\mathcal{E}_{max}$(TNSA) $\simeq 2.4$. For the MNA target, $\mathcal{E}_{max}$ drastically increases from $t \simeq 100$ fs and reaches $\mathcal{E}_{max} \simeq 600$ MeV at $t = 250$ fs through the two acceleration phases, i.e., the run-up and main-drive phases. Furthermore, the MNA protons continue to be accelerated even after the laser illumination ceases, reaching $\mathcal{E}_{max} \simeq 900$ MeV at the end ($t > 800$ fs). The phase with this additional acceleration effect is hereafter referred to as the "afterburner phase", which is one of the outstanding features of the MNA target. In contrast to the MNA target, the proton acceleration of the H-rod and the foil targets practically ceases, when the laser illumination is completed at $t \sim 250$ fs.

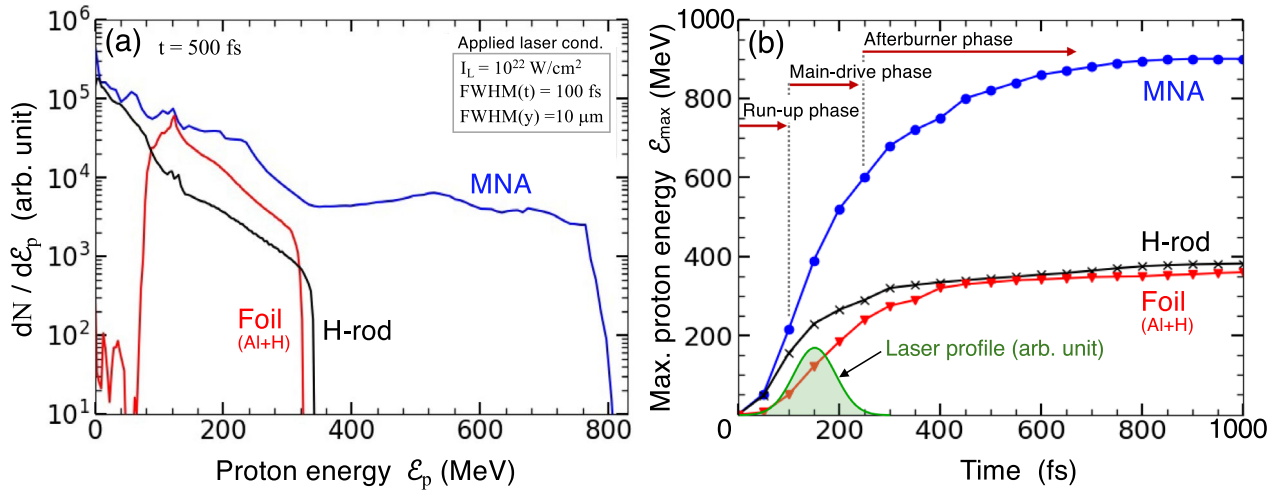

**Fig. 4.** (**a**) The proton energy spectra for the MNA, the H-rod, and the foil targets. The H-rod (2μm-diameter) is just the bullet body taken out from the MNA target, while the foil target is composed of a 1μm-thick aluminum with 50nm-thick solid hydrogen layer over-coated on the rear surface. (**b**) The temporal evolution of the three cases of (**a**). For the MNA target, the maximum proton energy $\mathcal{E}_{\max}$ is read to be increasing at the rate of 160 MeV/100 fs even after the laser illumination is finished at $t \simeq 250$fs.

### Afterburner phase
The additional energy gain in the afterburner phase, $\Delta\mathcal{E}_{\max} \simeq 300$ MeV, substantially contributes to the final energy, $\mathcal{E}_{\max} \simeq 900$ MeV (Fig. 4). Below, we clarify the mechanism and present a simple analytical model to understand the underlying physics of the afterburner phase.

Figure 5 shows time-sequential snapshots of the electric field (upper row) and the proton density (lower row) in the afterburner phase, which is defined as the period, $t \geq 250$ fs, which is the practical end of the pulse when 99% of the entire laser energy has been invested (see the laser profile in Fig. 4b). The three timings ($t = 230, 250,$ and 270 fs) correspond to the early stage of the afterburner phase. Two different electric fields drive the protons along the $x$-axis (Fig. 5), i.e., the stationary and comoving electric fields. The stationary electric field, which is generated at the nozzle exit, is the main drive force. The comoving electric field contributes to the afterburner acceleration. In fact, the locations of the comoving electric field and the head protons are in phase, as indicated by the paired white dashed circles on the upper and lower rows. Indeed, the protons are further accelerated at the rate, 160 MeV/100 fs at $t = 250$ fs, as can be observed in Fig. 4b. The process of the proton acceleration in the afterburner phase is understood as an effective energy transfer from the thermal energy of hot electrons to the kinetic energy of protons via free expansion of the plasma. The afterburner effect is attributed to the nozzle structure, which well collimates the plasma jet composed of protons and electrons ejected from the nozzle.

Figure 6 shows the phase space plots at $t = 250$ fs and $t = 800$ fs. Comparing the initial positioning of the MNA target shows that the proton energies increase mainly after they are ejected from the nozzle through the main-drive and the afterburner phases. At $t = 250$ fs, the increase in proton momentum via the afterburner phase is still small, because the measured time is just at the beginning of the free expanding process, corresponding to $\mathcal{E}_{\max} \sim 600$ MeV (compare Fig. 4b). The momentum increase due to afterburner is observed more apparently in Fig. 6b, in which the thin green belt of the proton momentum substantially extends up to the high momentum region, $p_x/m_p c > 1.5$, corresponding to $\mathcal{E}_{\max} \sim 900$ MeV.

Here, we quantitatively approximate the additional increase $\Delta\mathcal{E}_{\max}$ through the afterburner phase on the basis of a self-similar solution in terms of a two-fluid (ion and electron) model[14]. This model describes a non-relativistic expansion of a finite plasma mass in a vacuum while fully considering the charge separation effects, and thus describing the electric field self-consistently. The characteristic ion energy of such an expanding plasma with an isothermal electron temperature $T_e$ is given by[14,54]

$$\mathcal{E}_{i0} = 2ZT_e \ln(R/R_0), \quad (1)$$

where $R_0$ and $R = R(t)$ are the system sizes at the initial state ($t = 0$) and and at time $t$, respectively. The electron temperature generated by an intense laser is approximately given by the formula[13], $T_e(\text{MeV}) \simeq 44\sqrt{I_{L22}\lambda_{L\mu}^2}$, where $I_{L22}$ and $\lambda_{L\mu}$ are the laser intensity and the laser wavelength in units of $10^{22}$ W/cm$^2$ and $1\mu$m, respectively. When the laser illumination ceases and the free plasma expansion begins, the system size is roughly estimated to be $R \approx c_s \tau_L$, where $c_s = \sqrt{ZT_e/m_i}$ is the sound speed with $m_i$ being an ion mass under consideration.

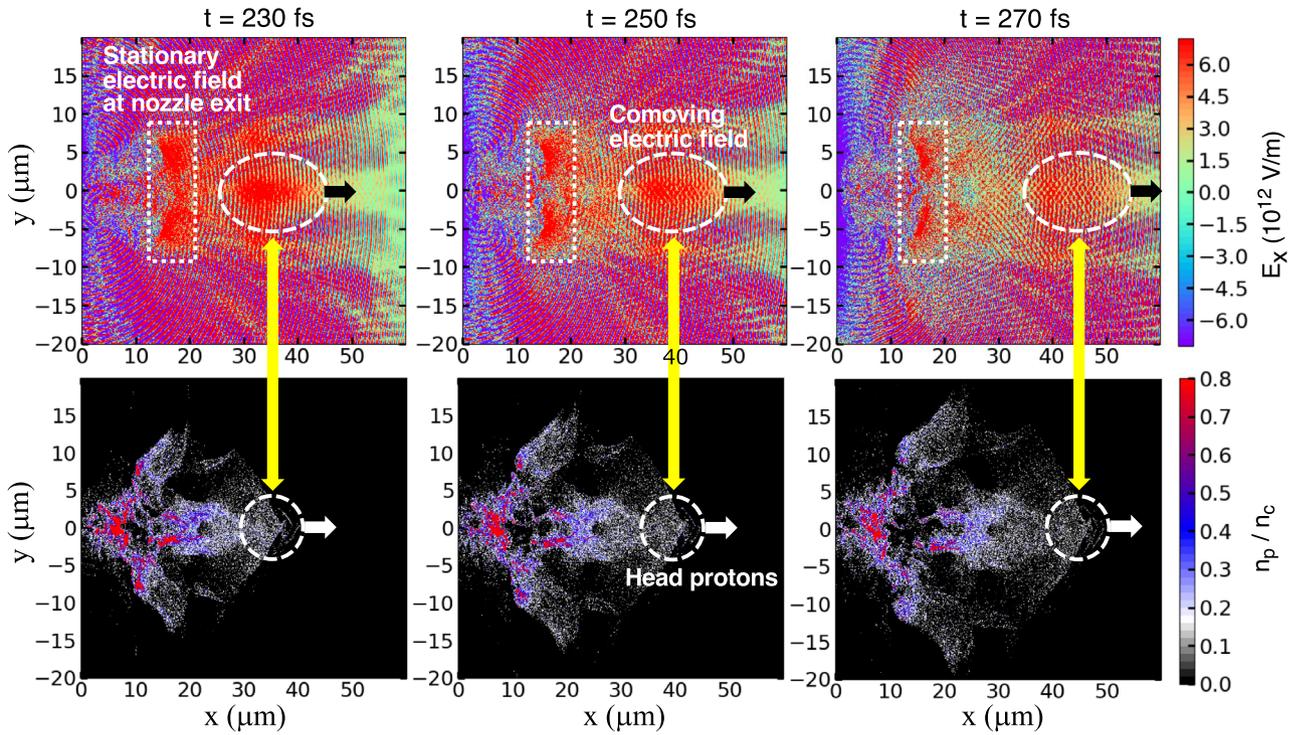

**Fig. 5.** Upper row: 2D profiles of the longitudinal electric fields $E_x$ (compare Fig. 1). Lower row: proton density profiles at different times, at sequential times, $t = 230, 250,$ and $270$ fs for the MNA target, which correspond to the early times of the afterburner phase; $n_c$ denotes the critical density. The head protons are observed to be continuously accelerated by the comoving electric field. Applied laser and target conditions are the same as in Fig. 3.

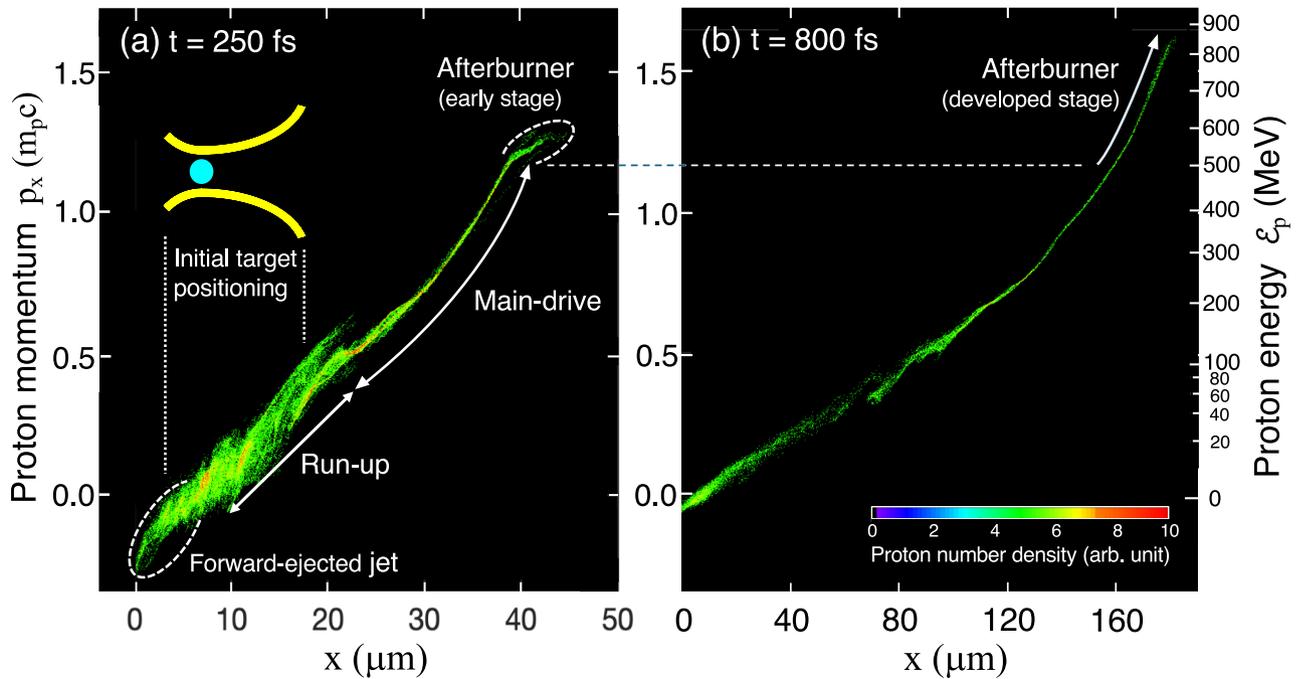

**Fig. 6.** The phase space plots at t = 250 fs and 800 fs, showing how the protons are accelerated through the three different phases. The proton energies increase mainly after they are ejected out of the nozzle through the main-drive and the afterburner phases.

Within the framework of the self-similar model, the normalized position of the ion front, $\xi_f = x_f/R$, is given in the form,

$$\xi_f^2 = W\left(\Lambda^2/2\right), \qquad (2)$$

where $\Lambda = R(t)/\lambda_D(t) = R_0/\lambda_{D0}$ is a dimensionless parameter with $\lambda_{D0} = \sqrt{T_e/4\pi n_{e0}e^2}$ being the Debye length at the initial electron density $n_{e0}$. The function, $W(x)$, is the inverse function of $x = W\exp(W)$, called Lambert function[55]; it behaves asymptotically as $W(x) \approx x$ for $x \ll 1$, while $W(x) \approx \ln(x/\ln x)$ for $x \gg 1$. Although the temporal constancy of $\Lambda$ is just the ansatz for the self-similar solution to exist, it turns out to be an acceptably practical assumption for plasma expansions under laser parameters in laboratory experiments. As a result, the self-similar solution approximately gives the additional energy gained through the afterburner phase via the free expansion into vacuum in the form,

$$\Delta\mathcal{E}_{\max} \sim \mathcal{E}_{i0}\xi_f^2 = 2ZT_e \ln\left(\frac{c_s\tau_L}{R_0}\right)W\left(\frac{\Lambda^2}{2}\right). \qquad (3)$$

Figure 7a plots the temporal evolution of $\mathcal{E}_{\max}$ of the MNA target for different applied laser intensities $I_L$. The curves are obtained under the same target structure employed in the previous figures. Figure 7b shows the corresponding $\mathcal{E}_{\max}$ at $t = 1$ ps as a reference (blue circles), and the increment of energies gained in the afterburner phase, $\Delta\mathcal{E}_{\max} = \mathcal{E}_{\max}(1\text{ps}) - \mathcal{E}_{\max}(250\text{fs})$ (orange circles). Both $\mathcal{E}_{\max}(1\text{ps})$ and $\Delta\mathcal{E}_{\max}$ are increasing functions of $I_L$. Moreover, the simple model, Eq. (3), which is depicted by the solid curve, well reproduces the simulation results, where the external parameters, $Z = 1$, $R_0 = 1\mu$m, $\tau_L = 100$ fs, $n_{e0} = 5 \times 10^{22}$ cm$^{-3}$, and $\lambda_L = 0.8\mu$m, are fixed. From the good agreement between the simulation results and the analytical model, it is inferred that the protons are continuously accelerated in the afterburner phase by absorbing the thermal energy of the hot electrons under the interplay between the charge-separated two fluids.

### Scaling of laser intensity

To explore the laser intensity scaling of MNA, additional simulations have been performed covering wide range of laser intensity $I_L$ from $1 \times 10^{20}$ to $5 \times 10^{22}$ W/cm$^2$. Figure 8 shows the summary of the simulation results for three different kinds of targets, i.e., MNA, H-rod, and foil targets. For the MNA target, the same set of the external parameters specifying the structure are employed as in Figs. 2 and 3. The MNA target is irradiated by either a plane pulse or a 10$\mu$m-wide (FWHM) Gaussian pulse. The other two cases, the H-rod ($D = 2\mu$m) and the foil target, are presented to compare with the MNA results. The H-rod target is just the same component embedded in the MNA target, which is irradiated by a 10$\mu$m-wide (FWHM) Gaussian pulse. The foil target is 24$\mu$m-wide and 1$\mu$m-thick solid aluminum coated with 50nm-thick solid hydrogen on the rear side[56], which is irradiated by a 10$\mu$m-wide (FWHM) Gaussian pulse.

It is noteworthy that the MNA target has a remarkably different $I_L$-scaling of $\mathcal{E}_{\max}$ from the other two types of targets. As depicted by the red and blue dashed lines in Fig. 8, the results for the MNA are well fitted by $\mathcal{E}_p \propto I_L^{0.79}$ in the range of $10^{20} \lesssim I_L \lesssim 10^{22}$, while those for the H-rod and the foil targets are well fitted by $\mathcal{E}_p \propto I_L^{0.5}$ as depicted by the gray and green lines. It should be noted that the absolute values for $\mathcal{E}_{\max}$ of the MNA target are obtained only under the fixed laser and target parameters and therefore they are not optimized. Here also note that $\mathcal{E}_{\max}$ of MNA target is found to further increase by 5 - 10% under periodic boundary condition along the y-axis (not presented in this paper). This suggests that a horizontally arrayed MNA target (standing side by side), for example, can provide easier circumstance in focal alignment and higher proton energies than a single MNA target.

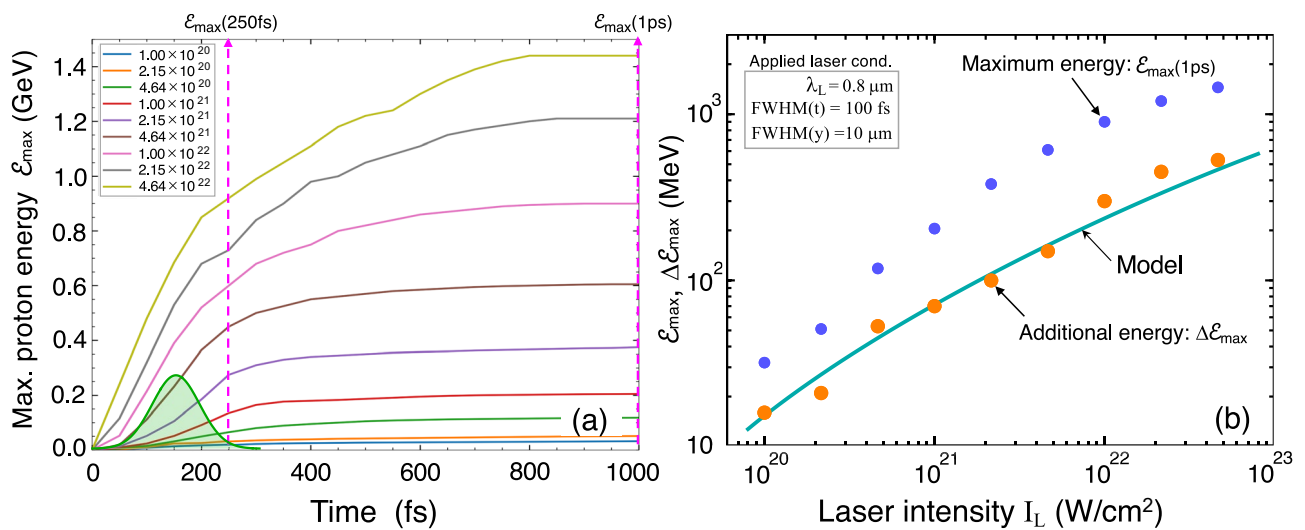

**Fig. 7.** (**a**) Temporal evolution of the maximum proton energies at different peak laser intensities (the upper-left inset in units of W/cm$^2$). The green Gaussian profile denotes the laser pulse. The time, $t = 250$ fs, is taken here as the beginning of the afterburner phase. (**b**) Maximum proton energies at the end, $\mathcal{E}_{\max}(1\text{ ps})$ (blue circles), and the additional energies gained in the afterburner phase, $\Delta\mathcal{E}_{\max} = \mathcal{E}_{\max}(1\text{ ps}) - \mathcal{E}_{\max}(250\text{ fs})$ (orange circles). The solid curve is obtained by a simple analytical model based on a self-similar analysis.

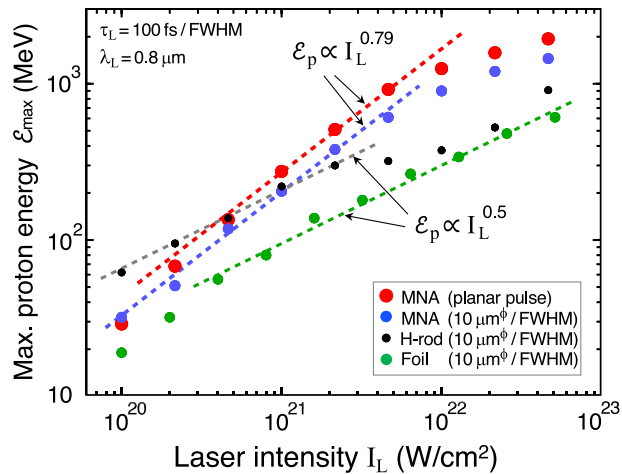

**Fig. 8.** Scaling of the maximum proton energy $\mathcal{E}_{\max}$ versus the laser intensity $I_L$, under the different laser and target conditions. The MNA target shows the stronger dependence on $I_L$ then the other two kinds of targets. The specific laser and targets parameters are the same as in Fig. 4.

The $I_L$-scaling of the foil target, $\mathcal{E}_p \propto I_L^{0.5}$, is widely accepted, for which the physical mechanism is understood based on the TNSA scheme[12] as follows. When an intense laser illuminates a foil target, absorbed laser energy first heats up the target to produce hot electrons, which fly to the rear side of the target to expand into vacuum. An electron sheath is then formed on the proton/vacuum boundary, where a strong electric field $E_f$ is generated. The electric field is estimated from the momentum balance, $\nabla p_e + n_e e E_f = 0$, where, $p_e$, $n_e$, and $e$ denote the electron pressure, the electron number density, and the electric charge, respectively. We postulate that the electrons isothermally expand into vacuum and that the resultant electric field accelerates the surface protons over the distance corresponding to the electron density scale $L_e$. As a result, the momentum balance reduces to $T_e/L_e + eE_f = 0$. The accelerated proton energy is then approximately given by $\mathcal{E}_p \sim eE_f L_e \sim T_e \sim I_L^{1/2}$, where the last relation $T_e \sim I_L^{1/2}$ is read from the ponderamorive scaling[13]. It should be noted that $L_e$ increases in time in a self-regulating manner during the plasma expansion. Therefore, in general, it does not coherently develop with the electron Debye length $\lambda_{De} = \sqrt{T_e/4\pi n_e e^2}$, i.e., $L_e/\lambda_{De} \neq$ const.

As shown in Fig. 8, the $I_L$-scaling of MNA is significantly different from those of the foil and the H-rod targets. Below we roughly estimate the scaling in terms of a simple argument. Suppose that an isolated matter with a scale length $R$ is illuminated by an intense laser. As was observed in Figs. 3, 4 and 5, a long-life and wide-stretched electrostatic field is formed in MNA targets. This is because a certain amount of net electron charge $Q$ are stripped off from the MNA target after the laser heating. If a hot electron with a thermal kinetic energy $T_e$ can barely escape to infinity, the energy balance between the electrostatic potential energy and the kinetic energy reads $eQ/R \sim T_e$, which then approximately gives the electric field on the target surface as $E_f \sim Q/R^2 \sim T_e/eR$. As confirmed in Figs. 3, 4, 5 and 6, protons are dominantly accelerated in the vicinity of the nozzle exit, where low density plasma composed of ions and electrons are flowing out. Here, we assume that protons are effectively accelerated over the distance corresponding to the Debye length $\lambda_{De}$. Consequently, the proton energy is estimated to be $\mathcal{E}_p \sim eE_f \lambda_{De} \sim T_e^{3/2} \sim I_L^{3/4}$, which is close to the scaling $\mathcal{E}_p \propto I_L^{0.79}$ obtained by the simulation. Note that, due to disintegration of Al-nozzle, the $I_L$-scaling begins to deviate from $\mathcal{E}_p \propto I_L^{0.79}$ to make its inclination smaller for the high intensity region, $I_L \gtrsim 10^{22}$ W/cm$^2$, as seen in Fig. 7. Also note that, at the low intensities ($I_L \lesssim 10^{21}$ W/cm$^2$), the MNA target shows inferior performance in the $I_L$-scaling compared with the H-rod. The physical reason for the inferior performance is inferred that a strong electric field in the nozzle skirt cannot be generated under such low laser intensities.

The same scaling for the foil target (TNSA scheme), $\mathcal{E}_p \propto I_L^{0.5}$, applies to the cylindrical H-rod target, as long as $D \gg L_e$ holds, when the system can be practically regarded as planar geometry at the lower laser intensities ($I_L \lesssim 6 \times 10^{21}$ W/cm$^2$ in Fig. 8). However, at the higher laser intensities, the $I_L$-dependence of the H-rod shows significantly different behavior, i.e., the plotted points in Fig. 8 once nearly level off ($6 \times 10^{21} \lesssim I_L \lesssim 3 \times 10^{22}$ W/cm$^2$) and then increase again with $I_L$, for the following physical reason. The higher $I_L$, the longer $\lambda_{De}$, and consequently the proton expansion becomes more like Coulomb explosion, in which the laser electric field is much stronger than the Coulomb field of the cluster proton core. In Coulomb explosions, most of electrons are instantaneously blown off in the time scale of $D/c$ with $c$ being the speed of light, while keeping the cold protons at their initial positions. The maximum proton energy $\mathcal{E}_p$ is then a function only of the initial diameter $D$ and the density. In other words, $\mathcal{E}_p$ does not depend on $I_L$ beyond a certain critical value, as long as the protons expand in a cylindrically symmetric manner. This cylindrical symmetry is broken at even higher laser intensities ($I_L \gtrsim 3 \times 10^{22}$ W/cm$^2$), where a strong shock wave is driven to transmit in the H-rod target to boost the proton acceleration as a result[57–59].

### Laser pulse width and energy conversion efficiency

Although we have so far fixed the laser pulse width to be $\tau_L = 100$ fs, it is important to examine the $\tau_L$-dependence on the performance of proton acceleration. Figure 9 shows the maximum proton energy $\mathcal{E}_{max}$ and the laser-to-proton conversion efficiency $\eta_c$ as a function of $\tau_L$ under different applied laser intensities $I_L$. Except for $\tau_L$, all the target and laser conditions are the same as in Fig. 8, corresponding to the blue circles. For $I_L \lesssim 5 \times 10^{21}$ W/cm$^2$, $\mathcal{E}_{max}$ weakly increases with $\tau_L$. On the other hand, for $I_L \gtrsim 10^{22}$ W/cm$^2$, $\mathcal{E}_{max}$ has a peak around $\tau_L \sim 20$ fs, which is a favorable pulse width from a viewpoint of the energy conversion efficiency. In other words, for $5 \times 10^{21} \lesssim I_L(\text{W/cm}^2) \lesssim 1 \times 10^{22}$, $\tau_L \approx 20$ fs is an optimum pulse width, because such a laser pulse can lead to a good performance with a reasonable compromise between $\eta_c$ and $\mathcal{E}_{max}$. Moreover, if one employs the specific parameters, $I_L = 10^{22}$ W/cm$^2$ and $\tau_L = 20$ fs for the laser spot size of 10μm × 10μm, just for example, the required laser power and laser energy are $P_L = 10$ PW and $E_L = 200$ J, respectively. For further and detailed estimates of MNA performance, three-dimensional optimization is to be conducted. It should be noted that, as can be inferred from Fig. 9, the conversion efficiencies for $\tau_L \lesssim 25$ fs are likely comparable with those obtained by the TNSA scheme, which are experimentally reported to be $\eta_c \lesssim 3 - 4$ %.[60,61]

### Angular divergence

Angular divergence is one of the most critical issues when assessing proton beam quality from the application point of view. Figure 10 shows a summary on the angular distributions of proton beams, $dN/d\theta$, obtained by the PIC simulations at different laser intensities, where the angle $\theta$ is measured with respect to the x-axis. The coherent results for $\mathcal{E}_{max}$ are given in Fig. 8 as the blue circles. On the vertical plane in Fig. 10, the angular divergences (FWHM), $\Delta\theta(I_L)$, are plotted as red squares. While $\Delta\theta$ depends on the specific value of $I_L$, they distribute well within the range, $5° \lesssim \Delta\theta \lesssim 25°$. The average over all the laser intensities is found to be $\Delta\theta \simeq 16°$. Just as a reference, the comparison between the MNA and the foil target based on the TNSA scheme reveals $\Delta\theta(\text{MNA}) = 18°$ while $\Delta\theta(\text{TNSA}) = 23°$ at a specified laser intensity $I_L = 1 \times 10^{22}$ W/cm$^2$. As a whole, MNA can provide proton beams with relatively reduced angular divergences owing to the nozzle structure, which works to effectively collimate the proton fluxes.

### Practical target structure with rod/nozzle contact

In the previous sections, we have considered such a prototype that has finite gaps between the H-rod and the nozzle frames, i.e., $\Delta = H_2 - D \neq 0$. From a practical point of view of target fabrication, we here examine the acceleration performance for three different cases, in which the nozzle frames are attached to the H-rod, causing $\Delta = 0$. As shown in the inset of Fig. 11, we then consider the four kinds of targets, i.e., (A) Prototype ($D = 2$μm and $\Delta = 0.8$μm), (B) Narrow neck (the nozzle frames are simply attached to the $D = 2$μm rod), (C) Large rod ($D = 2.8$μm), and (D) Elliptic rod (the lengths of semi-minor and semi-major axes are $D_x = 0.7^{-1}$μm and $D_y = 2.8$μm, respectively, to keep the cross-sectional area the same as that of the prototype, i.e., $D_x D_y = D^2$). Otherwise, the same nozzle and laser parameters are applied to all the four cases.

The four curves in Fig. 11 show the corresponding proton energy spectra at $t = 800$ fs, which are more or less similar to each other. This is because the H-rod is almost transparent to the hot electrons with a temperature $\gtrsim$ a few tens of MeV, and therefore the existence of material-filled gap does not significantly affect the acceleration dynamics of protons. It is notable, however, that the maximum proton energies for cases C and D, $\mathcal{E}_{max} \approx 1$ GeV, are even higher by a factor of $\sim 10\%$ than $\mathcal{E}_{max} \approx 0.9$ GeV for case A. Besides, the laser-to-proton energy conversion efficiency of case D is roughly twice higher than that of case A, which can be accounted by the mass ratio of the H-rods between cases A and D. Thus, it is left as a crucial task to optimize the laser and target conditions to maximize the acceleration performance of MNA.

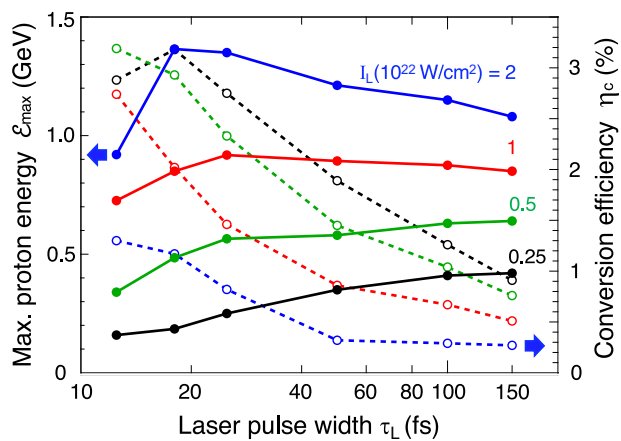

**Fig. 9.** Maximum proton energy $\mathcal{E}_{max}$ and the laser-to-proton conversion efficiency $\eta_c$ measured at $t = 1$ ps as a function of the laser pulse width $\tau_L$ under different applied laser intensities $I_L$. All the target and laser conditions are the same as those for the blue circles in Fig. 8.

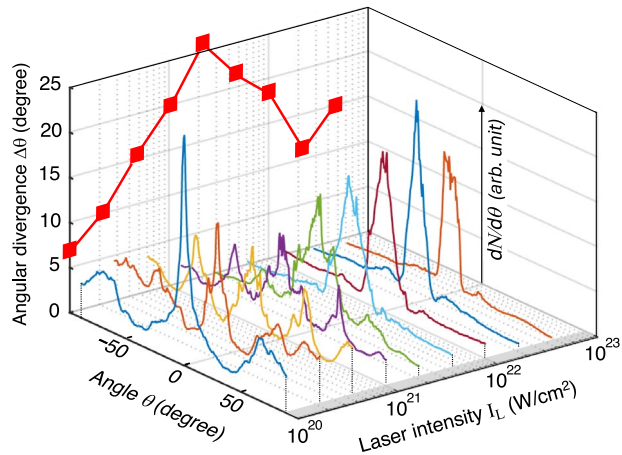

**Fig. 10**. Summary on the angular distributions of the proton beams, $dN/d\theta$, at different laser intensities, where the angle $\theta$ is measured with respect to the x-axis. The coherent results for the maximum proton energy $\mathcal{E}_p$ are given in Fig. 8 as the blue circles. The red squares on the vertical plane denote the angular divergences (FWHM), $\Delta\theta(I_L)$, obtained from the individual curves of $dN/d\theta$.

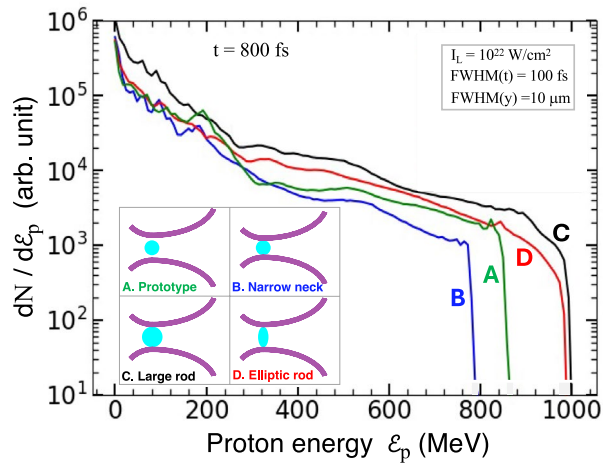

**Fig. 11**. Proton energy spectra for the different four cases shown in the inset. The nozzle frames are attached to the H-rod except for the prototype (case A), which corresponds to the simulation results for the MNA target given in Figs. 4 , 5 and 6.

## Summary

The novel ion acceleration scheme, micronozzle acceleration (MNA), has been proposed. In MNA, the micronozzle functions like a power lens, such that laser energy absorbed by the nozzle leads to the generation of strong electric fields to accelerate the protons contained in the center-located H-rod. The protons undergo continuous acceleration through the three different phases, i.e., the run-up, main drive, and the afterburner phases. In particular, from the comparison between the simple analytical model and the simulation results, it has turned out that the protons are continuously accelerated by absorbing the thermal energy of the hot electrons through plasma expansion in the afterburner phase.

In MNA, a maximum proton energy $\mathcal{E}_{max} \simeq 1$GeV is obtained at an applied laser intensity $I_L \simeq 1 \times 10^{22}$ W/cm$^2$. The scaling of $\mathcal{E}_{max}(I_L)$ for the MNA is obtained as $\mathcal{E}_{max} \propto I_L^{0.79}$, which contrasts with that for the TNSA scheme, $\mathcal{E}_{max} \propto I_L^{0.5}$. While the maximum proton energy $\mathcal{E}_{max}$ only weakly depends on the pulse width $\tau_L$, the maximum conversion efficiency $\eta_c \sim 3\%$ can be obtained with $\tau_L \lesssim 25$ fs. Further optimizations with respect to the target structure and the laser conditions are anticipated to lower the threshold of necessary laser intensity to achieve GeV-class proton beams. It should be noted that a proof-of-principle experiment for MNA can be designed even under moderate laser conditions.

### Data availability
The datasets used and/or analysed during the current study are available from the corresponding author on reasonable request.

### References
1. Cowan, T. E. et al. Ultralow emittance, multi-MeV proton beams from a laser virtual-cathode plasma accelerator. *Phys. Rev. Lett.* **92**, 204801 (2019).
2. Romagnani, L. et al. Dynamics of electric fields driving the laser acceleration of multi-MeV protons. *Phys. Rev. Lett.* **95**, 195001 (2005).
3. Borghesi, M. et al. Measurement of highly transient electrical charging following high-intensity laser-solid interaction. *Appl. Phys. Lett.* **82**, 1529 (2003).
4. Krushelnick, K. et al. Energetic proton production from relativistic laser interaction with high density plasmas. *Phys. Plasmas* **7**, 2055 (2000).
5. Pukhov, A. et al. Three-dimensional simulations of ion acceleration from a foil irradiated by a short-pulse laser. *Phys. Rev. Lett.* **86**, 3562 (2001).
6. Murakami, M. et al. Generation of high-quality mega-electron volt proton beams with intense-laser-driven nanotube accelerator. *Appl. Phys. Lett.* **102**, 163161 (2013).
7. Patel, P. et al. Isochoric heating of solid-density matter with an ultrafast proton beam. *Phys. Rev. Lett.* **23**, 125004 (2003).
8. Roth, M. et al. Fast ignition by intense laser-accelerated proton beams. *Phys. Rev. Lett.* **86**, 436 (2001).
9. Krushelnick, K. et al. Ultrahigh-intensity laser-produced plasmas as a compact heavy ion injection source. *IEEE Trans. Plasma Sci.* **28**, 1184 (2000).
10. Bulanov, S. V. et al. Feasibility of using laser ion Accelerators in proton therapy. *Brief Commun.* **28**, 453 (2002).
11. Foukal, E. et al. Particle selection for laser-accelerated proton therapy feasibility study. *Med. Phys.* **30**, 1660 (2003).
12. Snavely, R. A. et al. Intense high-energy proton beams from petawatt-laser irradiation of solids. *Phys. Rev. Lett.* **85**, 2945 (2000).
13. Wilks, S. C. et al. Energetic proton generation in ultra-intense laser-solid interactions. *Phys. Plasmas* **8**, 542 (2001).
14. Murakami, M. et al. Self-similar expansion of finite-size non-quasi-neutral plasmas into vacuum: Relation to the problem of ion acceleration. *Phys. Plasmas* **13**, 012105 (2006).
15. Qiao, B. et al. Stable GeV ion-beam acceleration from thin foils by circularly polarized laser pulses. *Phys. Rev. Lett.* **102**, 145002 (2009).
16. Henig, A. et al. Radiation-pressure acceleration of ion beams driven by circularly polarized laser pulses. *Phys. Rev. Lett.* **103**, 245003 (2009).
17. Gonzalez-Izquierdo, B. et al. Radiation pressure-driven plasma surface dynamics in ultra-intense laser pulse interactions with ultra-thin foils. *Appl. Sci.* **8**, 336 (2018).
18. Silva, L. O. et al. Proton shock acceleration in laser-plasma interactions. *Phys. Rev. Lett.* **92**, 015002 (2004).
19. Fiuza, F. et al. Laser-driven shock acceleration of monoenergetic ion beams. *Phys. Rev. Lett.* **109**, 215001 (2012).
20. Yin, L. et al. GeV laser ion acceleration from ultrathin targets: The laser break-out afterburner. *Laser Part. Beams* **24**, 291 (2006).
21. Yin, L. et al. Break-out afterburner ion acceleration in the longer laser pulse length regime. *Phys. Plasmas* **18**, 063103 (2011).
22. Kuznetsov, A. V. et al. Efficiency of ion acceleration by a relativistically strong laser pulse in an underdense plasma. *Plasma Phys. Rep.* **27**, 211 (2001).
23. Bulanov, S. V. & Esirkepov, TZh. Comment on "Collimated multi-MeV ion beams from high-intensity laser interactions with underdense plasma". *Phys. Rev. Lett.* **98**, 049503 (2007).
24. Nishihara, K. et al. High energy ions generated by laser driven Coulomb explosion of cluster. *Nucl. Instrum. Methods Phys. Res. A* **464**, 98 (2001).
25. Murakami, M. et al. Efficient generation of quasimonoenergetic ions by Coulomb explosions of optimized nanostructured clusters. *Phys. Plasmas* **16**, 103108 (2009).
26. Nakamura, T. et al. Coulomb implosion mechanism of negative ion acceleration in laser plasmas. *Phys. Rev. A* **373**, 2584 (2009).
27. Wagner, F. et al. Simultaneous observation of angularly separated laser-driven proton beams accelerated via two different mechanisms. *Phys. Rev. A* **22**, 063110 (2015).
28. Kluge, T. et al. High proton energies from cone targets: Electron acceleration mechanisms. *New J. Phys.* **14**, 023038 (2012).
29. Honrubia, J. J. et al. On intense proton beam generation and transport in hollow cones. *Matter Radiat. Extremes* **2**, 28 (2017).
30. Rusby, D. R. et al. High proton energies from cone targets: Electron acceleration mechanisms. *Matter Radiat. Extremes* **103**, 053207 (2021).
31. Sgattoni, A. et al. Laser ion acceleration using a solid target coupled with a low-density layer. *Phys. Rev. E* **85**, 036405 (2012).
32. Yang, Y. C. et al. Proton acceleration from laser interaction with a complex double-layer plasma target. *Phys. Plasmas* **25**, 123107 (2018).
33. Pugachev, L. P. et al. Acceleration of electrons under the action of petawatt-class laser pulses onto foam targets. *Nucl. Instrum. Methods Phys. Res. A* **829**, 88 (2016).
34. Prencipe, I. et al. Acceleration of electrons under the action of petawatt-class laser pulses onto foam targets. *Plasma Phys. Control. Fusion* **58**, 034019 (2016).
35. Chatterjee, G. et al. Macroscopic transport of mega-ampere electron currents in aligned carbon-nanotube arrays. *Phys. Rev. Lett.* **108**, 235005 (2012).
36. Jiang, S. et al. Microengineering laser plasma interactions at relativistic intensities. *Phys. Rev. Lett.* **116**, 085002 (2016).
37. Khaghani, D. et al. Enhancing laser-driven proton acceleration by using micro-pillar arrays at high drive energy. *Sci. Rep.* **7**, 11366 (2017).
38. Gozhev, D. A. et al. Laser-triggered stochastic volumetric heating of sub-microwire array target. *Plasma Phys. Control. Fusion* **37**, 100856 (2020).
39. Murakami, M. et al. Generation of ultrahigh field by micro-bubble implosion. *Sci. Rep.* **8**, 7537 (2018).
40. Murakami, M. et al. Relativistic proton emission from ultrahigh-energy-density nanosphere generated by microbubble implosion. *Phys. Plasmas* **26**, 043112 (2019).
41. Ji, L. L. et al. Towards manipulating relativistic laser pulses with micro-tube plasma lenses. *Sci. Rep.* **6**, 23256 (2016).
42. Zou, D. B. et al. Efficient generation of 100 MeV ions from ultrashort E21Wcm-2 laser pulse interaction with a waveguide target. *Nucl. Fusion* **59**, 066034 (2019).

43. Murakami, M. et al. Generation of megaTesla magnetic fields by intense-laser-driven microtube implosions. *Sci. Rep.* **10**, 16653 (2020).
44. Liu, J. L. et al. Two-stage acceleration of protons from relativistic laser-solid interaction. *Phys. Rev. ST Accel. Beams* **15**, 101301 (2012).
45. Liu, J. L. et al. Stable laser-produced quasimonoenergetic proton beams from interactive laser and target shaping. *Phys. Rev. ST Accel. Beams* **16**, 121301 (2013).
46. Bake, M.A. et al. Energetic protons from an ultraintense laser interacting with a symmetric parabolic concave target. *Phys. Plasmas* **20**, 033112 (2013).
47. Kaymak, V. et al. Boosted acceleration of protons by tailored ultra-thin foil targets. *Sci. Rep.* **9**, 18672 (2013).
48. Bulanov, S. S. et al. Generation of GeV protons from 1 PW laser interaction with near critical density targets. *Phys. Plasmas* **17**, 043105 (2010).
49. Zhang, Z. M. et al. High-density highly collimated monoenergetic GeV ions from interaction of ultraintense short laser pulse with foil in plasma. *Phys. Plasmas* **17**, 043110 (2010).
50. Zou, D. B. et al. Enhanced laser-radiation-pressure-driven proton acceleration by moving focusing electric-fields in a foil-in-cone target. *Phys. Plasmas* **22**, 023109 (2015).
51. Rahman, O. et al. High-quality GeV proton beam generation from multiple-laser interaction with double-layer target. *Phys. Plasmas* **28**, 053106 (2021).
52. Lezhnin, K. V. & Bulanov, S. V. Laser ion acceleration from tailored solid targets with micron-scale channels. *Phys. Rev. Res.* **4**, 033248 (2022).
53. Arber, T. D. et al. Contemporary particle-in-cell approach to laser-plasma modeling. *Plasma Phys. Control. Fusion* **57**, 113001 (2015).
54. Murakami, M. et al. Ion energy spectrum of expanding laser-plasma with limited mass. *Phys. Plasmas* **12**, 062706 (2005).
55. Corless, M. et al. On the Lambert W Function. *Adv. Comput. Math.* **5**, 329 (1996).
56. Balusu, D. et al. Ion acceleration from aluminum foil coated with a gold nanolayer irradiated by ultrashort laser pulses. *Phys. Plasmas* **57**, 113001 (2024).
57. Matsui, R. et al. Quasimonoenergetic proton bunch acceleration driven by hemispherically converging collisionless shock in a hydrogen cluster coupled with relativistically induced transparency. *Phys. Rev. A* **31**, 013107 (2024).
58. Matsui, R. et al. Dynamics of the boundary layer created by the explosion of a dense object in an ambient dilute gas triggered by a high power laser. *Phys. Rev. Lett.* **122**, 014804 (2019).
59. Jinno, S. et al. Laser-driven multi-MeV high-purity proton acceleration via anisotropic ambipolar expansion of micron-scale hydrogen clusters. *Sci. Rep.* **12**, 16753 (2022).
60. Nishiuchi, M. et al. Efficient production of a collimated MeV proton beam from a polyimide target driven by an intense femtosecond laser pulse. *Phys. Plasmas* **15**, 053104 (2008).
61. Green, J. S. et al. High efficiency proton beam generation through target thickness control in femtosecond laser-plasma interactions. *Appl. Phys. Lett.* **104**, 214101 (2014).

## Acknowledgements
This work was supported by the Japan Society for the Promotion of Science (JSPS) and the Kansai Electric Power Company, Incorporated (KEPCO). All the computations were performed by using the supercomputer SQUID under the support of the D3 Center, Osaka University.

## Author contributions
M.M. conceived the study and wrote the paper. Designing and performing the PIC simulations are contributed mainly by D.B. The two authors, M.M. and D.B., contributed equally to this work, and are therefore the co-first authors. S.M. and Y.M. contributed to the PIC simulations. B.R. contributed to many useful discussions and suggestions. All authors reviewed the whole work, and approved the manuscript.

## Declarations

### Competing interests
The authors declare no competing interests.

## Additional information
**Correspondence** and requests for materials should be addressed to M.M.

**Reprints and permissions information** is available at www.nature.com/reprints.

**Publisher's note** Springer Nature remains neutral with regard to jurisdictional claims in published maps and institutional affiliations.